\documentclass{article}

\usepackage{arxiv}
\usepackage[utf8]{inputenc} 
\usepackage[T1]{fontenc}    
\usepackage{hyperref}      
\usepackage{url}            
\usepackage{booktabs}      
\usepackage{amsfonts}       
\usepackage{nicefrac}       
\usepackage{microtype}      
\usepackage{graphicx}

\title{An Overview of Agent-based Traffic Simulators}

\author{
  Johannes Nguyen, Thomas Farrenkopf,\\
  \textbf{Michael Guckert}\\ 
  Kompetenzzentrum für Informationstechnologie\\ 
  Technische Hochschule Mittelhessen\\ 
  61169 Friedberg, Germany\\
  \texttt{\{johannes.nguyen,thomas.farrenkopf,} \\
  \texttt{michael.guckert\}@mnd.thm.de} \\
  \And
  Simon T. Powers, Neil Urquhart \\
  School of Computing\\
  Edinburgh Napier University\\
  EH10 5DT, Edinburgh, United Kingdom\\
  \texttt{\{s.powers,n.urquhart\}@napier.ac.uk} \\
}

\begin{document}
\maketitle

\begin{abstract}
Individual traffic significantly contributes to climate change and environmental degradation. Therefore, innovation in sustainable mobility is gaining importance as it helps to reduce environmental pollution. However, effects of new ideas in mobility are difficult to estimate in advance and strongly depend on the individual traffic participants. The application of agent technology is particularly promising as it focuses on modelling heterogeneous individual preferences and behaviours. In this paper, we show how agent-based models are particularly suitable to address three pressing research topics in mobility: 1. Social dilemmas in resource utilisation; 2. Digital connectivity; and 3. New forms of mobility. We then explain how the features of several agent-based simulators are suitable for addressing these topics. We assess the capability of simulators to model individual travel behaviour, discussing implemented features and identifying gaps in functionality that we consider important.
\end{abstract}


\section{Introduction}
\label{sec:Intro}
Over the last decades, transportation and personal mobility have repeatedly faced radical changes. Driven by technological innovation and changing societal demands, traffic and transportation have evolved into complex systems. For example, Intelligent Transportation Systems (ITS) use advanced information technology to improve traffic management. Central traffic control systems are deployed to provide real-time information on road closures, parking space availability, etc. in order to minimise avoidable problems in traffic. In many countries, ITS are already in use with the main objectives to increase general traffic safety and to make more efficient use of the existing infrastructure. Computer-based simulations can be used to plan and assess the effects of new policies in advance, and provide decision support for transport planners and authorities. State of the art research on traffic simulation has shown a growing interest in the application of multi-agent models. Multi-agent models are implementations of \textit{Decentralised Artificial Intelligence} \cite{ferber1999multi}. They are an established means for the \textit{construction of synthetic worlds} which can be used to simulate and analyse interactions of complex systems \cite{ferber1999multi}. \cite{zheng2013primer} provide a description of common structures found in agent platforms that are designed for the simulation of traffic. Agents are \textit{closed computer systems that are situated in some environment, and that are capable of autonomous action in this environment in order to meet their designed objectives} \cite{wooldridge1997agent}. This autonomous and goal-driven behaviour of intelligent software agents makes agent models particularly suitable for the representation of individuals in road traffic. For example, travellers can be modelled as agents that interact and perceive information about their environment through sensors, allowing for implementation of decentralised knowledge and thus autonomous behaviour based on situational conditions. This approach to modelling individuals is a key distinction of multi-agent approaches from other types of simulation models e.g. cellular automata~\cite{clarke2014cellular}. \newline

As ideas on traffic simulation date back to the 1970s \cite{poeck1976anwendung,axhausen1989simulating} a variety of computer-based simulators has been developed. Due to the broad spectrum of traffic simulators that have emerged over the years, interdisciplinary end-users working on specific research questions are faced with the issue of finding an appropriate simulation environment. \cite{jakob2013modular} describe the phenomenon that in many cases instead of exploiting the potential of already existing simulators, researchers have implemented their own research specific applications. This may be the result of not having sufficient overview on the set of available simulators and their features which requires a considerable amount of in-depth research. In this paper, we reflect on current research topics in mobility and provide examples of simulators with related case studies. The paper specifically addresses the group of researchers that are studying individuals and their reciprocal effects on traffic to help them get an impression of the broad range of simulators and their capabilities to model individual travel behaviour.

This paper is organised as follows: The next section provides an overview of the diversity of traffic simulators and presents scope and areas of application of simulation models included in our review. We give examples of simulators with related case studies for each area of application. Following this, a review of simulators with regard to their background, system architecture and modelling capabilities is given. The review is primarily based on publications and publicly available discussions of expert communities. After that, we reflect on current state of implementation for modelling of individuals and discuss missing functionality that can help to improve research into simulation of individuals and their behaviour. The paper ends with a discussion of future steps that can help close these gaps. 

\section{Perspectives of Comparison}
\label{sec:dimensionsofcomparison}
A search performed in September 2021 across three common publication databases: Google Scholar, ACM Digital Library and IEEE Xplore; delivered an overview of available simulators. 
We searched for peer-reviewed papers by keywords (1. Traffic Simulation; 2. Agent-based Traffic Simulation; 3. Multi-agent traffic simulation) contained in the title, abstract, and the main body of the papers. The first 30 research papers from each database and each keyword were included in a backward search to identify simulators that are considered related work by the authors of the publications. Furthermore, we also looked at some of the previous review papers (\cite{algers1997review,pursula1999simulation,passos2011towards}) as well as forum discussions\footnote{e.g. \url{https://www.researchgate.net/post/What-is-the-best-agent-based-traffic-simulation-tool} - (access on 13/05/2020)} from the research community in order to complete the search. With regard to the selection of papers included in this study, we only considered papers written in English and removed duplicate and irrelevant papers. We only looked at road traffic and excluded papers that focused on maritime or air traffic. In the case that different papers relate to the same simulator we used the most cited paper. Our search has produced a number of simulators that can be categorised based on their modelling approach. In the literature, traffic simulation models are divided into following four levels of detail \cite{passos2011towards,lopez2018microscopic}:

{\begin{enumerate}
  \item \textit{Macroscopic simulations} focus on traffic flow modelling based on high-level mathematical models. This type of simulation can be used for the analysis of wide-area systems in which no detailed modelling is required, e.g. the simulation of motorway traffic. Given the low level of detail, macroscopic simulations are relatively fast and require less computing power. 
  \item \textit{Microscopic simulations} focus on modelling individual entities based on a high level of detail. Possible entities include travellers, vehicles, traffic lights, etc. This type of simulation is often used for the analysis of urban traffic. It is possible to analyse both macroscopic and microscopic aspects (e.g. traffic lights algorithm, multimodal traffic) of the system. Consequently, microscopic simulations may result in longer computing times.
  \item \textit{Mesoscopic simulations} are a mixture of macroscopic and microscopic simulation models. Traffic entities are modelled at a higher level of detail than macroscopic approaches, however, interaction and behaviour of the individuals appear to be less detailed.
  \item \textit{Nanoscopic simulations} are even more detailed than microscopic approaches. This type of simulation is applied in the field of autonomous driving, in which internal functions of the vehicles such as gear shifting or vehicle vision have to be examined.  
\end{enumerate}
}

Agent models can be positioned as microscopic simulations that can also be used for research purposes with a higher level of detail (mesoscopic and macroscopic). 
The level of detail determines which aspects of the transport system are covered. Such differences are also reflected in the data required for modelling. The use of real-world data should increase the realism and accuracy of simulations. However, researchers need to be aware about the purpose of their simulation and choose a simulation model that supports the required level of detail for dealing with their research objectives. Going into more detail than necessary can make a simulation model complex and also requires more input data.
If we consider macroscopic, microscopic and nanoscopic simulations, they all have two fundamental elements within the problem scenario that must be defined by the input data:

\begin{itemize}
    \item \textbf{Demand}: The demand element defines the requirement for travel and thus the resulting traffic volume between locations. This can be modelled using either activity- or trip-based approaches. Depending on the selected modelling approach different input data are required. For example, activity-based approaches use information from census and behaviour surveys to generate daily activity schedules of individuals and thus creates the need to travel. In contrast, trip-based approaches make use of \textit{origin-destination} (OD) matrices which require no information on the daily schedules of individuals and thus allows for more abstract representation of traffic. However, trip-based approaches can also consider different levels of detail. At a macroscopic level this may be modelled through distributions of vehicles moving between larger areas, e.g. the number of vehicles per hour moving between a group of towns. Such information may come from traffic surveys or census data (e.g. giving the number of daily commuters between two towns). As we consider microscopic simulation, it becomes necessary to differentiate between individual vehicles. Rather than moving between two towns, demand may be in the form of specific journeys from an address to another address for a specific reason (e.g. commuting or shopping). Within mesoscopic simulation we might simulate journeys from a general location to a specific address, for instance commuter journeys that begin from a town, but travel to a specific employers' address. In order to simulate at the microscopic levels, we move from a high level OD matrix to a more detailed OD matrix, with entries for specific addresses. As we begin to specify demand through specific journeys use of travel diaries and census information allows us to learn the travel habits of individuals. Nanoscopic simulation often focus on a smaller geographical area, demand may be represented by those journeys that are completely within the simulation as well as those that either pass through the simulation or only start/end within the area. Demand is likely to be specified as individual journeys, once again best specified using census or travel diary data.
    \item \textbf{Infrastructure}: The infrastructure element comprises a representation of the road network. At a fundamental level, the road network comprises a graph of nodes and arcs that represent junctions and roads respectively. The amount of detail required at the macroscopic level is minimal possibly denoting that a route between two towns exists and its capacity/travelling time, possibly only taking into account trunk routes.  When using simulations for which greater levels of detail are required (e.g. microscopic and nanoscopic) it becomes necessary to include lower capacity roads and intermediate junctions in the road  graph. At the microscopic level, the graph will need to contain information such as lane capacities, and junction types. At this level the difference made by features such as traffic signals, turn restrictions or lane closures may radically affect the outcome of the simulation. OpenStreetMap (OSM) \cite{haklay2008openstreetmap} can provide a detailed source of road network data that can be applied at most levels of simulation.
\end{itemize}

In addition, the selection of algorithms significantly influences the options and limitations of the underlying simulation models. In this work, a distinction is made between the following categories: \textit{fully agent-based, featuring agent technology} and \textit{not agent-based}. We consider a simulator \textit{fully agent-based} when key concepts of the simulation (e.g. travellers, vehicles) are fully implemented as intelligent software agents. This leverages the individual perspective in the modelling that comes with capabilities for interaction as well as autonomous and goal-driven behaviour. Simulators that use agent technology to extend alternative approaches by agent capabilities for specific aspects of the simulation are referred to as simulators that \textit{feature agent technology}. Depending on the research objective general purpose platforms such as NetLogo \cite{wilensky1999center} provide basic agent functionalities that can be used implement lightweight experiments. In this paper, we focus on simulators that are designed for simulation of large-scale scenarios as these are most relevant to implement real-world case studies. Our literature search has produces the following list of simulators for each category of simulation models:

{
\begin{itemize}
  \item \textit{Fully Agent-Based}: MATSim  \cite{horni2016multi}, ITSUMO \cite{bazzan2010itsumo}, MovSim \cite{treiber2010open}, MASCAT \cite{gueriau2016assess}, MATISSE \cite{torabi2018matisse}, POLARIS \cite{auld2016polaris}, AgentPolis \cite{jakob2013modular}, OPUS \cite{waddell2006opus}, MOSAIIC \cite{czura2015mosaiic}, MARS \cite{weyl2018MARS}, SimMobility \cite{adnan2016simmobility}, SITRAS \cite{hidas1998sitras}, ArchiSim \cite{champion2001behavioral}, SEMSim (CityMOS) \cite{xu2012semsim}, JTSS \cite{tao2009extensible}, Megaffic + XAXIS \cite{osogami2012research}, SD-Sim \cite{dumbuya2002agent}, SM4T \cite{zargayouna2014simulating}, VCTS \cite{chaurasia2010virtual}, SIMTUR \cite{10.5555/2338776.2338779}, MUST \cite{10.5555/2484920.2485243}, CAMiCS \cite{10.5555/1400549.1400667}, OpEMCSS \cite{clymer2002simulation}, DEFACTO \cite{schurr2009defacto}, MAGE \cite{banos2007simulating}, CityScape \cite{ion2015agent}, BAE Systems \cite{handford2011agent}, AITSPS \cite{zhou2010design}, SeSAm \cite{klugl2006sesam}, IMAGES \cite{yoo2009images}, Mobiliti \cite{chan2018mobiliti}, CUPSS \cite{1393646}, KLMTS1.0 \cite{chen2008study}, CARLA \cite{dosovitskiy2017carla}, AgentStudio \cite{10.5555/1400549.1400563}, ILUTE \cite{salvini2005ilute}, SIMULACRA \cite{batty2013simulacra}, TransWorld \cite{wang2010parallel} 
   \item \textit{Featuring Agent-Technology}:  ATSim \cite{chu2011atsim}, FastTrans \cite{thulasidasan2009designing}
  \item \textit{Not Agent-Based}: TRANSIMS \cite{texas1999early}, SUMO \cite{krajzewicz2002sumo}, OpenTraffic \cite{miska2011opentraffic,tamminga2014open}, CONTRAM \cite{taylor2003contram}, PTV VISSIM/VISUM \cite{fellendorf1994vissim}, GETRAM/AIMSUN \cite{barcelo2005dynamic}, PARAMICS \cite{cameron1996paramics}, MITSIM \cite{yang2000simulation}, FreeSim \cite{miller2007freesim}, TSIS/CORSIM \cite{owen2000traffic}, VATSIM \cite{lei2001vatsim}, DRACULA \cite{liu2010traffic}, RENAISSANCE \cite{wang2006renaissance}, SimTraffic \cite{sorenson2000practical}, DynaMIT \cite{ben1998dynamit}, DYNASMART \cite{mahmassani1993network}, MITSIMLab \cite{yang1996microscopic}, CUBE Voyager \cite{bentley_systems_2021}, PELOPS \cite{wallentowitz1999effects}, TransModeler \cite{balakrishna2009large}, Dynameq \cite{mahut2010traffic}, CORFLO \cite{lieu1992corflo}, PACSIM \cite{cornelis2002pacsim}, SIMSCRIPT II.5. \cite{bernhard2000traffic}, CTSP \cite{elci1982city}, CityMob \cite{martinez2008citymob}, VanetMobiSim \cite{harri2006vanetmobisim}, FIVIS \cite{schulzyk2007bicycle}, THOREAU \cite{wang1995enhanced}, GENIVI \cite{wang2018enhanced}, SLX \cite{henriksen2000slx}, SALT \cite{song2018statistical}, SIM-ENG \cite{766447}, KAIST \cite{kwon2001kaist}, UMTSM \cite{4959152}, SES/MB \cite{537860}, SISTM \cite{hardman1996motorway}, INTERGRATION \cite{van1996integration}, MATDYMO \cite{choi2006development}, TRANSYT \cite{byrne1982handbook}
\end{itemize}
}

As our objective is to address modelling of individuals, there is a primary focus on the approaches of the first and the second category. \cite{moller2019future} have identified trending subjects in mobility for which a considerable amount of research and investment is currently focused. Based on these subjects, we consider three areas of application (\textit{Social dilemmas in resource Utilisation, Digital Connectivity and New Forms of Mobility}) for which we give examples of simulators that have been used to research issues related to this domain. We are aware that areas of applications are closely connected and therefore may be overlapping. Hence, simulators mentioned as an example do not have to be used exclusively for the mentioned area of application, but can be particularly helpful.   
The simulators are studied with regard to three key aspects. We present general information (background, programming language, license, etc.) that is relevant to the selection of simulators and give an overview of the system architecture, describing basic functionalities of sub-components. Furthermore, we discuss implemented features for modelling individuals with regard to their area of application.\newline

Modelling of individuals and their travel-related behaviour depends on the simulated level of detail which is closely linked to the considered time perspective of the simulation. \textit{Long-term} aspects for example refer to decisions about workplace and residency whereas \textit{short-term} decisions involve movements on a micro scale such as spontaneous interactions, lane changing, or acceleration and braking. \textit{Mid-term} behaviour are in between and consider pre-journey planning such as route choice or selection of travel modes. The mid-term perspective distinguishes two types of approaches to modelling travel demand. This can be trip-based or activity-based. In trip-based approaches travel demand is modelled using OD-matrices (origin-destination) that can be based on static values or probability distributions. Alternatively, trip-based approaches can also be modelled using LSP (location-specific probabilities) usually resulting in travellers moving in space with no route specification. Instead, locations are assigned a pair of probabilities for the number of travellers starting as well as stopping at the location. In contrast to this, activity-based demand modelling for example produces a set of activities (e.g. working in the office, going to the gym, going grocery shopping) for each traveller, thus creating the need to travel. In this case, OD-matrices are a consequence of generated activity schedules. Considering this, we review modelling capabilities of simulators with regard to these aspects.

\section{Review}
\label{sec:review}
As described in the previous section, we focus our review  on agent-based approaches as our objective is to address the issue of modelling individuals. We concentrate on three areas of application: 1. Social Dilemmas in Resource Utilisation; 2. Digital Connectivity; 3. New Forms of Mobility. Due to the number of available simulators it is not possible to review all of them within the scope of this paper. Therefore, for each application we will look at three examples of simulators that have been used to model issues related to this domain.

\subsection{Social Dilemmas in Resource Utilisation}
This application domain considers the issues arising from transport infrastructure inherently being a shared resource used by many individuals, but not owned by any one of them. This means that use of transport infrastructure by one individual often creates negative externalities that affect other individuals, e.g. congestion and pollution \cite{samuelson1993tragedy}. Better public transport and shared mobility services are intended to relieve the traffic load on roads, while electrification of vehicles is seen as a means to reduce exhaust fumes and environmental pollution. This creates new questions as to what effects will be achieved in the short term as well as in the long term. For example, E-mobility inevitably leads to a change in energy consumption that requires efficient planning of available resources. In this paper, we will look at MATSim, POLARIS and SimMobility as examples for agent-based simulators that have already been used to simulate issues in this context. Other simulators that also fall into this category include: SEMSim (CityMOS), Megaffic + XAXIS, MUST, CAMiCS, DEFACTO, CityScape, BAE Systems, SeSAm, Mobiliti, MARS, MOSAIIC, OPUS, ILUTE, SIMULACRA 

\subsubsection{MATSim}
\label{sec:matsim}
MATSim is an agent-based software framework implemented in Java and licensed under GPLv2 or later. The project started in 2004 at ETH Zurich and is currently being developed in collaboration with TU Berlin and CNRS Lyon. The framework has a general focus and is designed for the simulation of large-scale transportation scenarios. Hence, a particular effort was made for efficient computational processing and parallelisation \cite{dobler2011design,charypar2008efficient}. MATSim has been used in particular to simulate energy demand planning in transportation \cite{novosel2015agent}. \newline

\begin{figure}[hbp]
	\centering
	\includegraphics[width=0.9\textwidth]{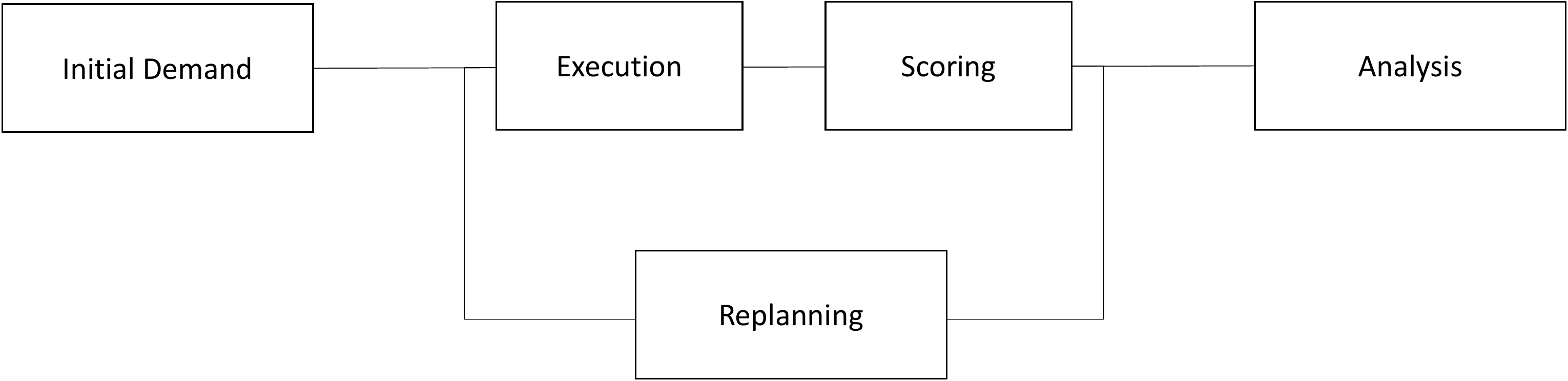}
	\caption{MATSim - Architecture.}
	\label {fig:matsim}
\end{figure}

The framework consists of five components for \textit{Initial Demand, Execution, Scoring, Replanning} and \textit{Analysis} (see Figure \ref{fig:matsim}) \cite{horni2016multi}. Based on the modular approach, custom components can be implemented and integrated into MATsim in order to replace or to upgrade  provided default operations. The first component deals with modelling and generation of an initial agent population. Agents select and execute plans in the \textit{execution component}. The \textit{scoring component} calculates a score for every plan based on a given utility function. This score is an indicator for accomplished agent utility. The \textit{replanning component} uses a \textit{co-evolutionary algorithm} for optimising this utility. In contrast to an ordinary evolutionary algorithm that searches for a global optimum the co-evolutionary algorithm is applied to evolve the set of agent plans of the travellers. The simulation cycle (execution - scoring - replanning) repeats until MATSim reaches an equilibrium and agent scores stabilise. Finally, the output data of the simulation is being aggregated in the \textit{analysis component}. \newline

With regard to the modelling capabilities of the application, MATSim can be considered a mid-term simulator as scenarios are commonly modelled for single days \cite{horni2016multi}. However, there are some experiments that have demonstrated the simulation of multi-day scenarios \cite{horni2012matsim}. MATSim provides two options for generating an initial population of agents which can be random or based on user input. Census information is used in order to model every traveller explicitly.  The application provides a number of predefined parameters that can be configured. MATSim follows an activity-based approach for modelling travel demand. Survey data is used to generate various lists with activities that are assigned to the agents. It should be noted that travel demand changes with every iteration of the simulation as the simulation includes a replanning mechanism for rescheduling of activities.
Furthermore, agents possess a list of plans that contains different combinations of actions and choices. This includes choices not only about classical traffic properties such as routes and travel mode but also time scheduling. MATSim uses a discrete-choice model for implementing agent decisions \cite{horni2016multi}. Quantitative methods are used to determine probabilistic distributions for alternative actions. Agents select plans based on calculated scores from the scoring component. A higher score increases the probability of a plan to be chosen (see \cite{flotterod2016choice}). Given the level of detail considered in modelling of individuals, MATSim is suitable for simulating scenarios that analyse social dilemmas in resource utilisation based on the amount and types of traffic (activities and modal choices) that emerges in the system.

\subsubsection{POLARIS}
\label{sec:polaris}
POLARIS is an open-source agent-based software framework written in C++. The project was first published in 2013 (see \cite{Auld_2014}) and is currently maintained at Argonne National Laboratory. The motivation behind POLARIS was to combine different traffic-related modelling aspects into a single framework that otherwise require a number of separate standalone software applications. In~\cite{auld2016polaris}, the authors of POLARIS argue that transportation research has focused on these aspects only in an isolated manner. However, simulation of complex systems requires a combined method. Early attempts to integrate the isolated models into a unified system have shown that resulting solutions are either inflexible, non-modular or inefficient. Based on this, the authors describe a need for a unified solution that enables inter-operability between the isolated models. The POLARIS framework has been proposed to address this issue \cite{auld2016polaris}. 
POLARIS focuses on large-scale transportation scenarios and has been used to analyse energy consumption of vehicles in the city of Detroit comparing scenarios that include current and future vehicle technologies \cite{islam2017impact}. \newline

\begin{figure}[htbp]
	\centering
	\includegraphics[width=0.9\textwidth]{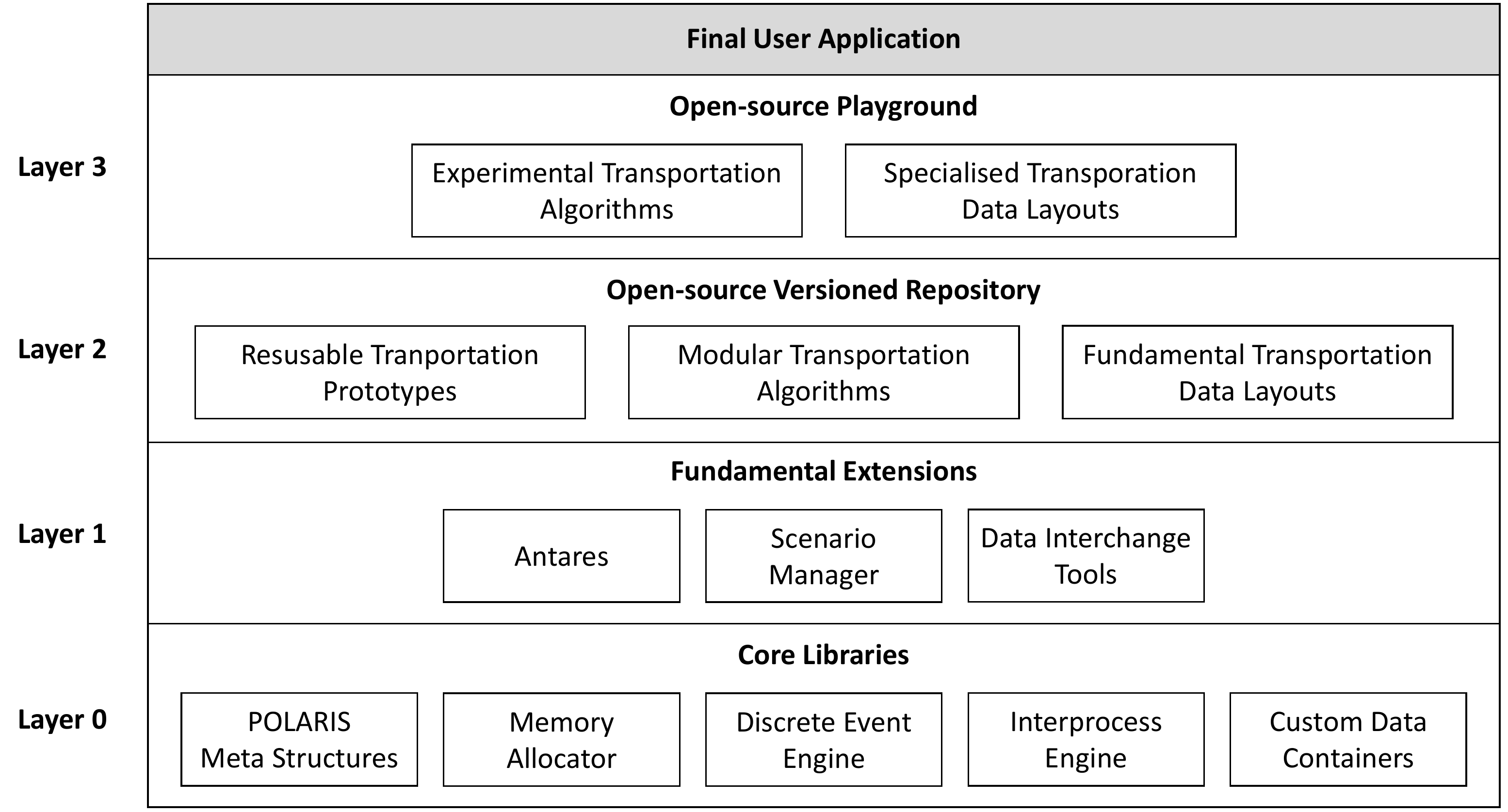}
	\caption{POLARIS - Architecture.}
	\label {fig:polaris}
\end{figure}

The framework provides a set of tools that can be used for the development, execution and review of a simulation model. The system architecture is structured using a layered approach (see Figure~\ref{fig:polaris}). Aspect-specific subcomponents are assigned to a layer depending on the level of modelling detail. This ensures abstract concepts which are commonly used across different variations of traffic simulation models to be less likely to change. Instead, users are supposed to make research-specific customisations on a more detailed level. This creates reusability of frequently used modelling aspects. Based on this, layer 0 is the most abstract layer of the POLARIS framework. Layer 0 contains a set of core libraries such as the discrete event engine which is responsible for handling  agents. Simulations are performed by executing a list of events. In layer 1, POLARIS contains a set of fundamental extensions. This includes components for 2D/3D visualisation (Antares) or data import/export services. Layer 2 is described as an \textit{open-source versioned repository}. In this repository, there is a set of model fragments that can be used for the implementation of custom simulation models. The provided model fragments are tested and chosen by universal applicability. Typical model fragments for example are reference implementations of well-established routing algorithms. Finally, layer 3 is described as the \textit{user playground}. In this layer, custom components can be included in order to extend the POLARIS framework with research-specific modelling aspects. Based on the provided elements from all layers, the user can build a custom application for agent-based traffic simulation.  \newline
 
With regard to modelling capabilities, POLARIS can be considered a mid-term simulator as travel decisions focus on mid-term aspects such as departure time, destination choice, route choices as well as planning and rescheduling of activities. Consequently, POLARIS uses an activity-based approach for modelling travel demand. This approach is based on an adjusted version of the ADAPTS \textit{(Agent-based Dynamic Activity Planning and Travel Scheduling)} model \cite{auld2009framework}. Originally, the ADAPTS model has been designed as a standalone application for simulating the occurrence of travel demand patterns that result from travel planning and scheduling processes. For integration into the POLARIS framework, the ADAPTS model has been reorganised in order to match the agent paradigm. This resulted in a separate \textit{activity planning} agent which as an extension to the traveller agent models the traveller's cognition of the activity planning process. This illustrates the applied structure for modelling other types of behaviour in POLARIS as a central traveller agent is composed of a set of subagents which each extend the traveller agent with cognitive capabilities for specific behavioural aspects. For example, these include agents for  perception, movement coordination or routing. In comparison to MATsim, this approach considers a more detailed modelling of individuals allowing for easier extension of short-term behaviour. This can be useful when energy consumption needs to be determined more precisely e.g. when simulating energy impact of acceleration and braking of autonomous vehicles to identify frequent nodes for charging stations.

\subsubsection{SimMobility}
\label{sec:simmobility}
SimMobility is a simulation platform written in C++ and published under an own open-source license. The project has related publications since 2015 and is currently developed at SMART (Singapore-MIT Alliance for Research and Technology) \cite{lu2015simmobility}. The simulator integrates a set of aspect-specific models relevant to the transportation domain that allows simulation on different time scales (short-, mid- and long-term) \cite{adnan2016simmobility}. For example, aspect-specific models include land-use, demographic movement or interactions related to transportation and communication. The platform focuses on modelling effects on traffic infrastructure, transportation services and the environment. This allows for the simulation of alternative planning options specifically with regard to technology, policies and investment. SimMobility has been used  to simulate the effects of new mobility services on the use of infrastructure \cite{marczuk2015autonomous}. \newline

\begin{figure}[htbp]
	\centering
	\includegraphics[width=0.7\textwidth]{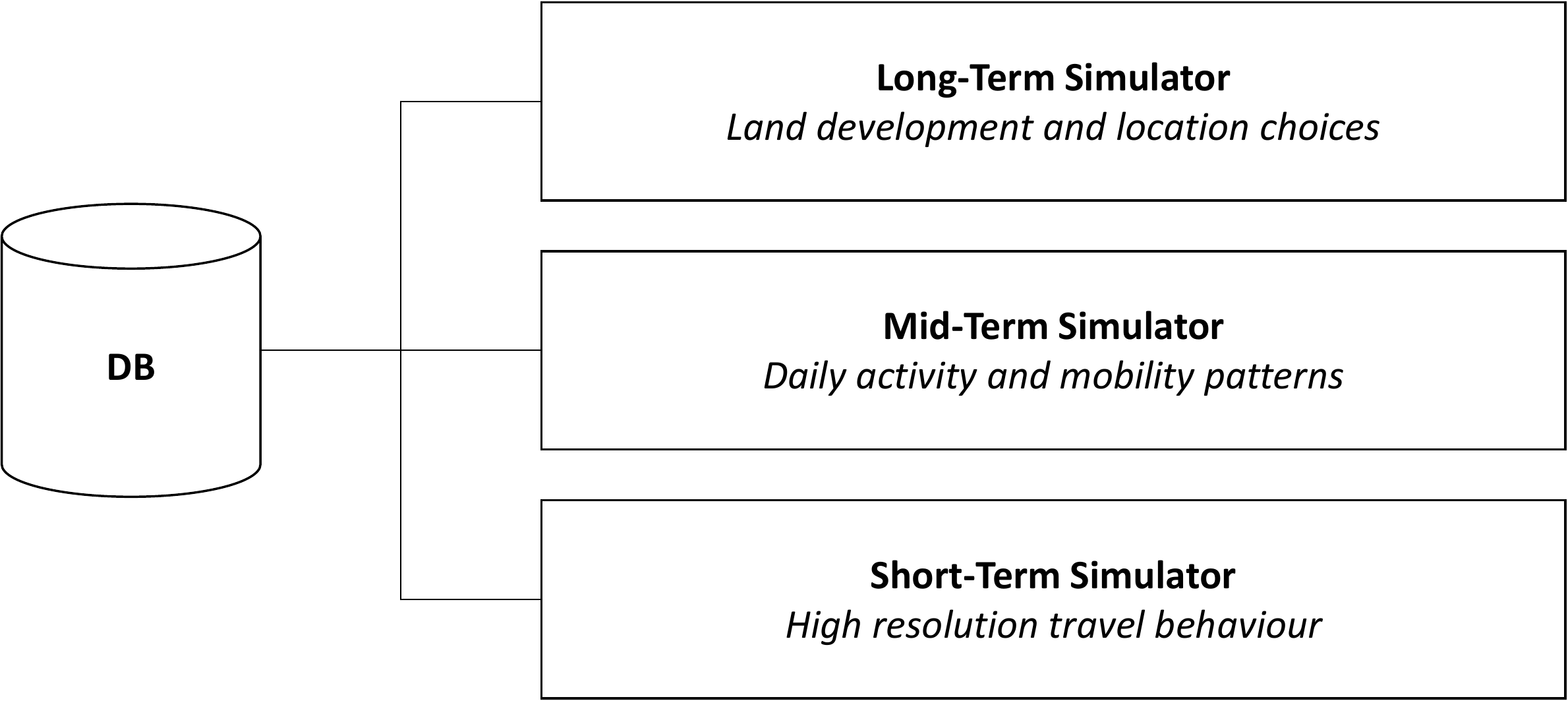}
	\caption{SimMobility - Architecture.}
	\label {fig:simmobility}
\end{figure}

The system architecture of SimMobility is structured in three components and follows a multi-level approach based on the time aspect. Each component simulates a different  perspective (see Figure~\ref{fig:simmobility}). The first component is the \textit{Long-term (LT)} simulator. This component deals with generating and updating the agent population. The LT simulator particularly simulates long-term aspects such as house location and car ownership, but also other long-term effects such as changes to the environment can be simulated in this module. The second component is described as the \textit{Mid-term (MT)} simulator \cite{lu2015simmobility}. This component is primarily designed for the simulation of agent behaviour in time scales of minutes and hours. This refers to \textit{high level} travel decisions such as route choice or modes of travel. The \textit{Short-term (ST)} simulator is the last component in the multi-level architecture which is a microsimulator based on MITSIM that has been extended with agent capabilities. A special characteristic of this architecture is that each component can be used as a standalone application. All simulators share the same database so that simulated individuals exist across all simulation levels simultaneously.

With regard to the modelling capabilities, SimMobility covers all time perspectives (long-, mid- and short-term) considered in this review and therefore is particularly flexible and powerful. Modelling aspects are distributed across the three subcomponents but are brought together into an individual using one database. SimMobility follows an activity-based approach for modelling travel demand \cite{adnan2016simmobility}. For each simulated day, the MT simulator generates a list of activities that include information on destination, departure time, route and mode choice. This approach has been integrated with methods of trip-based demand modelling as generated activities are aggregated to create origin-destination matrices that can be recalibrated. Agent decisions such as route choices are based on a probabilistic model which is similar to the MATSim approach \cite{azevedo2017simmobility}. The ST simulator also includes a mechanism that enables day-to-day agent learning to update the agent knowledge \cite{adnan2016simmobility}. Based on these modelling capabilities, SimMobility is probably the most flexible and powerful approach in this area of application with regard to modelling of individuals. Researchers that are uncertain about the required level of detail in modelling individual behaviour are able to easily adapt using this application. 

\subsection{Digital Connectivity}
The second type of application that we consider looks at the effect of the digital transformation on the mobility sector. For example, the use of digital traffic control systems (e.g. ITS), which can help to provide better driver experience for example by providing real-time information on parking and traffic jams, and also to improve transportation safety. In this paper, we look at the integration of SUMO and JADE as well as ITSUMO and MATISSE as examples of agent-based approaches that have been used for research on this type of simulation scenarios. Other simulators that also fall into this category include: SITRAS, ArchiSim, SM4T, SIMTUR, OpEMCSS, IMAGES, MASCAT, TransWorld 

\subsubsection{An Integration of SUMO and JADE}
\label{sec:sumo}
\textit{SUMO (}Simulation of Urban MObility) is a software framework for microscopic traffic simulation written in C++ that is licensed under EPL 2.0. A first version of the project was published in 2001 and created by the German Aerospace Center (DLR) \cite{krajzewicz2002sumo}. Since then, SUMO has been accepted by a wide community. The project was motivated by the necessity for an appropriate open-source solution as other projects which are now open-source, were difficult to obtain at that time \cite{horni2016multi}. Traffic applications were mainly used as black-boxes with no options to examine the underlying simulation model \cite{krajzewicz2002sumo}. Thus, researchers were restricted by the given parameterisation and modelling with no options to implement custom ideas. The SUMO approach is not agent-based but has been integrated with the \textit{Java Agent Development Framework (JADE)} (see \cite{bellifemine2005jade}) in order to make simulations compatible with recent agent technologies \cite{soares2013agent,azevedo2016jade}. JADE is an open-source software framework licensed under LGPLv2 that is used for the implementation of agent-based applications. This combination of SUMO and JADE has been used for simulating and assessing the effects of traffic control systems \cite{azevedo2016jade,timoteo2012using}. The following section on the system architecture focuses on the integration of SUMO and JADE. \newline

\begin{figure}[h]
	\centering
	\includegraphics[width=0.37\textwidth]{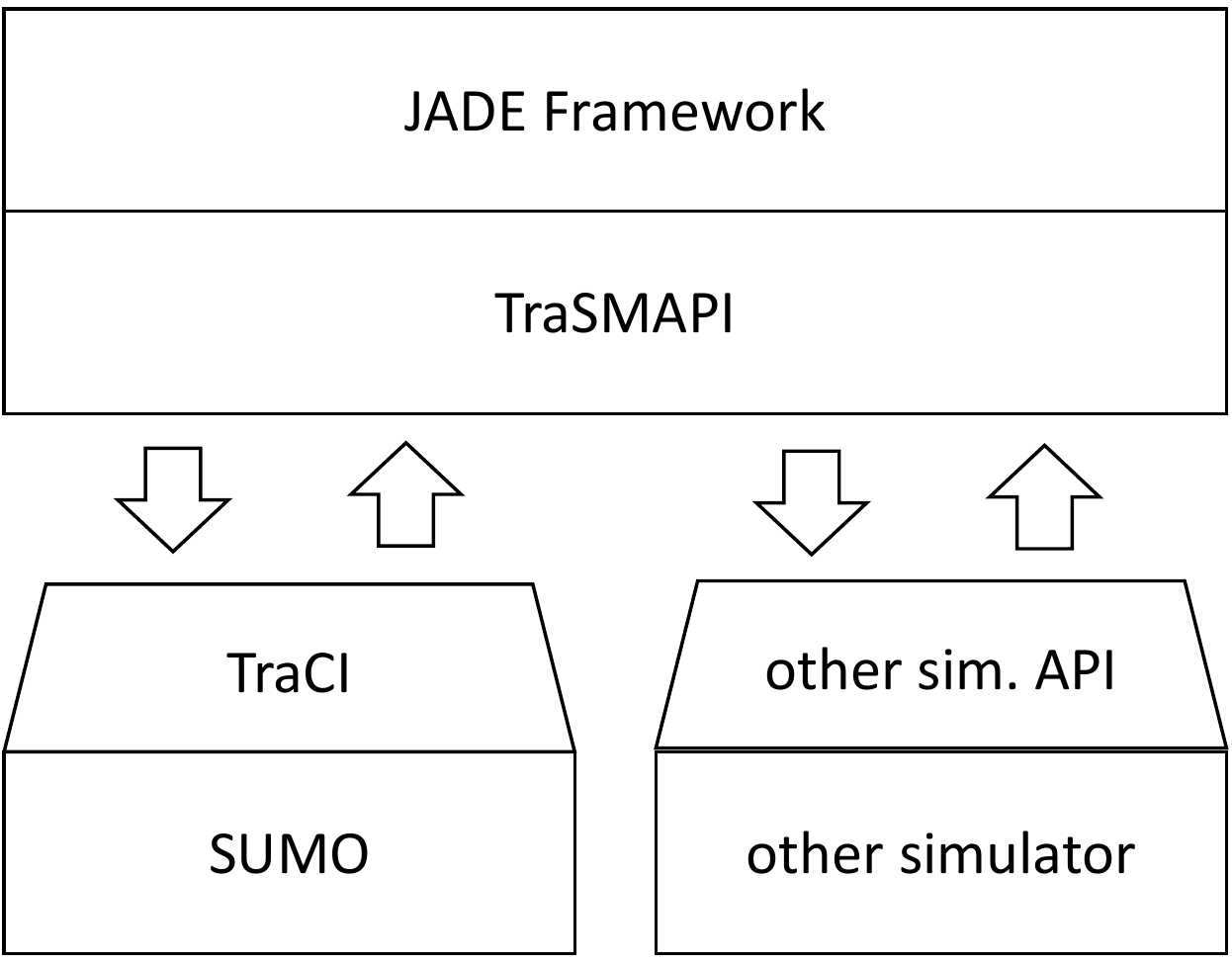}
	\caption{TraSMAPI - Architecture.}
	\label {fig:trasmapi}
\end{figure}

\cite{soares2013agent} have implemented a software connector that enables communication between the two software environments. This connector is referred to as \textit{TraSMAPI (Traffic Simulation Manager Application Programming Interface)}. From the SUMO perspective, the TraCI API is the central component for the integration of SUMO and JADE. TraSMAPI communicates with the TraCI API and acts as an intermediary. Although the project focuses on the integration of SUMO and JADE, TraSMAPI is abstracted to be able to handle various simulators besides SUMO (see Figure~\ref{fig:trasmapi}). This makes it possible to compare the results of different simulators. The combination of SUMO, JADE and TraSMAPI can therefore be termed as an \textit{Artificial Transportation System (ATS)} which is an extension of traditional modelling and simulation approaches with the ability to integrate different simulation models in a virtual environment \cite{wang2005framework}. \newline

\begin{figure}[htbp]
	\centering
	\includegraphics[width=0.7\textwidth]{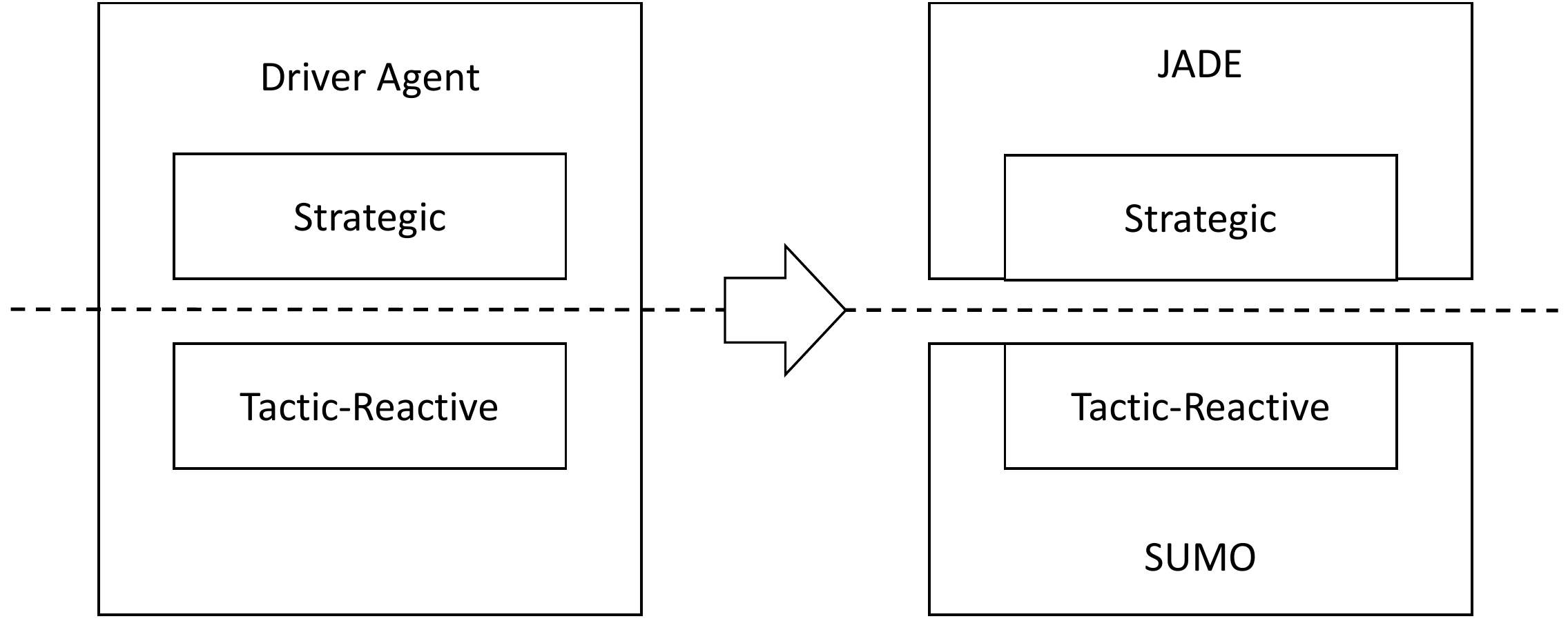}
	\caption{Delegate-Agent Concept.}
	\label {fig:delegate-agent-concept}
\end{figure}

With regard to the modelling capabilities, this approach is suitable for mid- and short-term simulations as modelling aspects include selection of travel modes but also micro-behaviour such as lane changing. JADE agents represent drivers that are linked to vehicles in SUMO. A separation of strategic and tactic-reactive agent behaviour has been implemented with two layers which is also referred to as the \textit{delegate-agent concept} \cite{wahle2002impact}. Basically, it can be understood as a separation of cognitive and reactive actions from the executing driving tasks \cite{soares2013agent}. The strategic layer deals with collection and processing of information from the surrounding environment. Based on this information the agent chooses its travel route, also in the strategic layer. In the tactic-reactive layer driving related behaviour such as acceleration, braking or lane changing is implemented. Based on the functional requirements of the two layers, the strategic layer was kept in JADE whereas the tactic-reactive layer was realised in SUMO (see Figure~\ref{fig:delegate-agent-concept}). 
The original SUMO package provides two options for demand modelling which can be trip-based using an origin-destination matrix  \cite{lopez2018microscopic} or using an activity-based approach.

Agent decisions are based on a probabilistic model but can be extended using the TraCI API. \cite{soares2013agent} have demonstrated the application of reinforcement learning techniques to model adaptable knowledge representation. Given the microscopic level of detail in modelling of individual behaviour, this application is suitable for simulating scenarios that analyse effects of traffic control policies on driving behaviour of individuals e.g. examining the perception of digital and analog traffic signs.

\subsubsection{ITSUMO}
\label{sec:itsumo}
ITSUMO (Intelligent Transportation System for Urban Mobility) is an open-source agent-based traffic simulator written in C++ and Java. The simulator was first presented in 2006 by UFRGS (Federal University of Rio Grande do Sul) and since then has been continuously refined and advanced \cite{da2004itsumo,bazzan2010itsumo}. Apart from the similarity in name, there is no direct link between ITSUMO and the previously described SUMO project. As the creators describe, ITSUMO was developed out of the lack of customising options in available simulation tools, as most of the existing solutions were developed for specific purposes. Other drawbacks described are for related simulation tools to not being fully agent-based, for them to be relying on strong simplifying assumptions, or deficiencies with regards to their demand planning options \cite{bazzan2010itsumo}. Thus, the ITSUMO approach is fully agent-based and aims at addressing the deficiencies mentioned above. ITSUMO has also been applied for the simulation of route choice scenarios. However, primary focus of the application is on traffic control. For example, ITSUMO has been used for testing traffic light algorithms \cite{rossetti2014advances,bazzan2010itsumo}.\newline

\begin{figure}[htbp]
	\centering
	\includegraphics[width=0.8\textwidth]{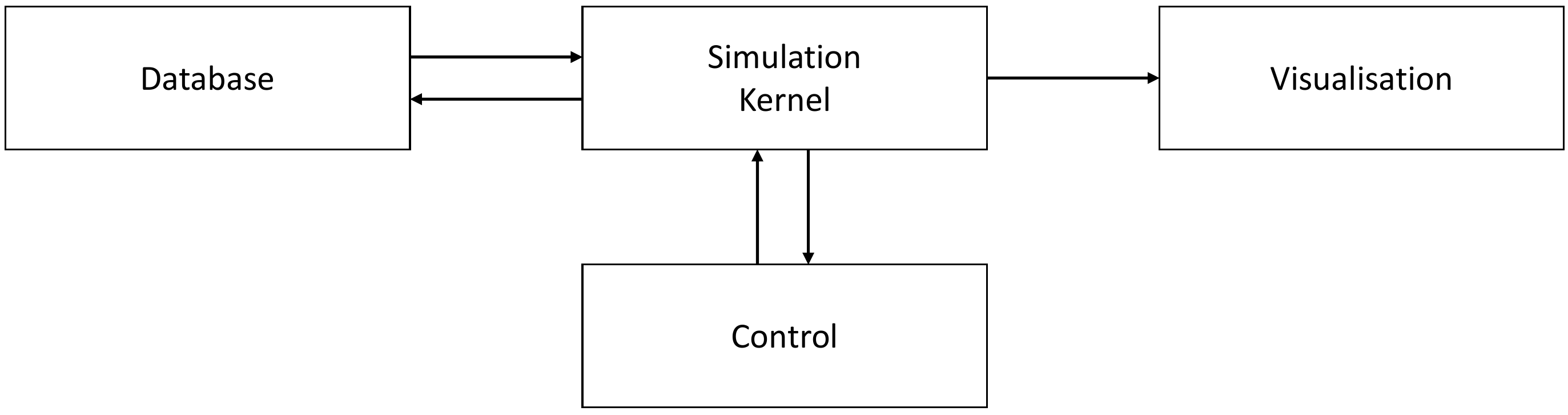}
	\caption{ITSUMO - Architecture.}
	\label {fig:itsumo}
\end{figure}

The system architecture is structured in five components \cite{bazzan2011extending,rossetti2014advances} (see Figure~\ref{fig:itsumo}). The first component is a \textit{database}. This database contains information about the geographic traffic network as well as other data used in the simulation (e.g. insertion rate of vehicles or origin and destination of the drivers). The second component is described as the \textit{simulation kernel}. This component accesses data stored in the database, executes the simulation and manages agent interaction. The system architecture also includes a separate \textit{control} component in which traffic-related control entities (e.g. traffic lights) are implemented. The control component passes information to the simulation kernel to provide instructions for simulated control entities. Finally, results of the simulation are \textit{output} in a separate component. For this, sensors and detectors are used during the simulation in order to collect relevant data such as travel times, average speed, etc. The output module provides two visualisation options for both, a microscopic and macroscopic view of the simulation. If the visualisation is not used, simulation data can also be output as files.\newline

With regard to the modelling capabilities, ITSUMO can be considered a mid-term simulator that focuses on control and assignment of travel demand. Therefore, travel decisions refer to the level of route choice as well as its spontaneous replanning. Agents can either replan at every intersection or in case of a delay during the journey. ITSUMO follows a trip-based approach for modelling travel demand. Travel demand can be modelled using an origin-destination (OD) matrix or by generating a synthetic demand using uniform probabilities for a set of locations (LSP). For each combination of origin and destination, vehicles are generated and a route is determined. The application is particularly suitable for simulations that deal with ITS as it provides specific interfaces for implementing control measures and the driver reactions that are related to them.

\subsubsection{MATISSE (DIVAs 4)}
\label{sec:matisse}
MATISSE is a large-scale agent-based simulation platform written in Java \cite{torabi2018matisse,mavs2015}. The simulator has been released by UTD MAVS (University of Texas at Dallas) for non-commercial use under GPLv3 using name \textit{DIVAs~4}. Early work related to the project has been published since 2004 during a time when only a few fully agent-based approaches existed \cite{Mili2004}. Within this set of fully agent-based simulation models, the creators of MATISSE criticised the lack of core agent mechanisms such as sensing, diverse communication types, etc. The project has been developed to overcome these deficiencies. MATISSE specialises in the simulation of scenarios related to traffic safety. \newline

\begin{figure}[htbp]
	\centering
	\includegraphics[width=0.6\textwidth]{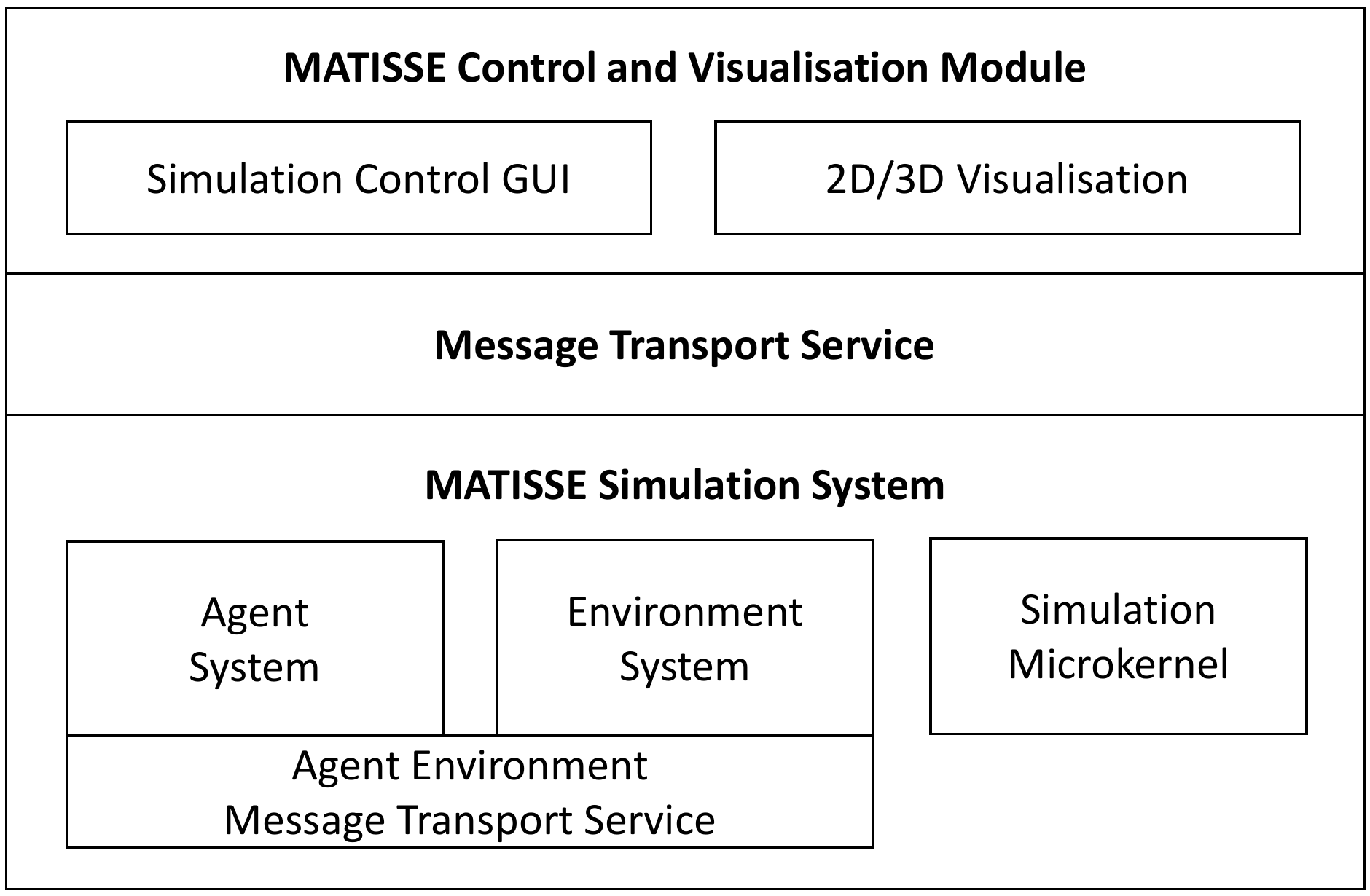}
	\caption{MATISSE - Architecture (Simplified).}
	\label {fig:matisse}
\end{figure}

The MATISSE architecture is structured in three layers (see Figure~\ref{fig:matisse}) \cite{torabi2018matisse}. The first layer is described as \textit{MATISSE Control and Visualisation Module}. It includes a control GUI for parameterisation and configuration of the simulation model. Furthermore, 2D/3D visualisation is implemented in this layer. Apart from this, there is a communication layer. This layer includes a \textit{Message Transport Service} that acts as a controller in order to enable communication between the user interface and the simulation system. The third layer \textit{MATISSE Simulation System} is the core element of the application. In this layer, calculations are performed in order to run the simulation. The layer is divided into three subsystems. The first subsystem is called \textit{Agent System}. This subsystem is responsible for the creation and control of various agents types (vehicles, traffic lights, etc.). The \textit{Agent-to-Agent Message Transport Service} handles agent communication during the simulation. The second subsystem is described as the \textit{Environment System}. This subsystem creates and controls additional simulation elements related to the traffic environment. This includes elements such as the traffic network. A separate \textit{Agent-Environment Message Transport Service} connects the environment system with the agent system. Finally, a third subsystem is the \textit{Simulation Microkernel}. This subsystem handles all tasks related to the simulation workflow. 

With regard to the modelling capabilities, MATISSE can be considered a mid- and short-term simulator as modelling aspects focus on driver behaviour. Similar to the ITSUMO approach, MATISSE also provides implementation for spontaneous replanning of route choices. Agent movement is based on car-following and lane-changing models, and it is even possible to model a virtual level of distraction that causes unpredicted traffic behaviour. The internal agent structure resembles a mental-level model from qualitative decision theory (see \cite{mccarthy1979ascribing}) which can be useful for modelling individuals. Furthermore, mental-level models provide a uniform basis for the comparison of agent behaviour which helps theoretical analysis \cite{brafman1997modeling}. MATISSE follows a trip-based approach for modelling travel demand using LSP. MATISSE uses a normal distribution or a user specified distribution in order to initialise agents for defined user entry and exit points. 
The application is particularly suitable for dealing with simulations on transportation safety and already provides a wide range of implementations for this area of application. The implemented mental-level structure of agents in MATISSE can be helpful for researchers that want to expand in their work on modelling and analysis of individual travel behaviour.

\subsection{New Forms of Mobility}
Ideas on improving the use of shared resources as well as increasing connectivity driven by technological innovation leads to new forms of mobility, which we consider as our final application area. This includes the deployment of new mobility services (e.g. ridesharing or -hailing) but also achievements in the field of autonomous driving. Mobility thus is influenced by diverse interactions between travellers and providers of mobility services, but also (autonomous) vehicles. In this paper, we look at AgentPolis, ATSim and MovSim as examples of agent-based applications that can help to work on interactions with new mobility services or coordination dynamics of autonomous driving. Other simulators that also fall into this category include: SD-Sim, VCTS, AITSPS, CARLA

\subsubsection{AgentPolis}
\label{sec:opentraffic}
AgentPolis is a fully agent-based software framework written in Java and licensed under GPLv3 \cite{jakob2012agentpolis,jakob2013modular}. The project was published in 2013 and created by AI Center FEE CTU (Czech Technical University in Prague). The creators noted that existing simulation approaches fail to implement the ability to model ad hoc interactions among the entities of the transport system as well as the spontaneous decision behaviour that is required for this form of interaction. However, current mobility services (e.g ridesharing) rely on frequent, ad hoc interactions between various entities of the transport system. Hence, AgentPolis focuses particularly on the simulation of interaction-rich transport systems. For example, the simulator has been used as a testbed for benchmarking on-demand mobility services \cite{vcerticky2014agent}. \newline

\begin{figure}[htbp]
	\centering
	\includegraphics[width=.55\textwidth]{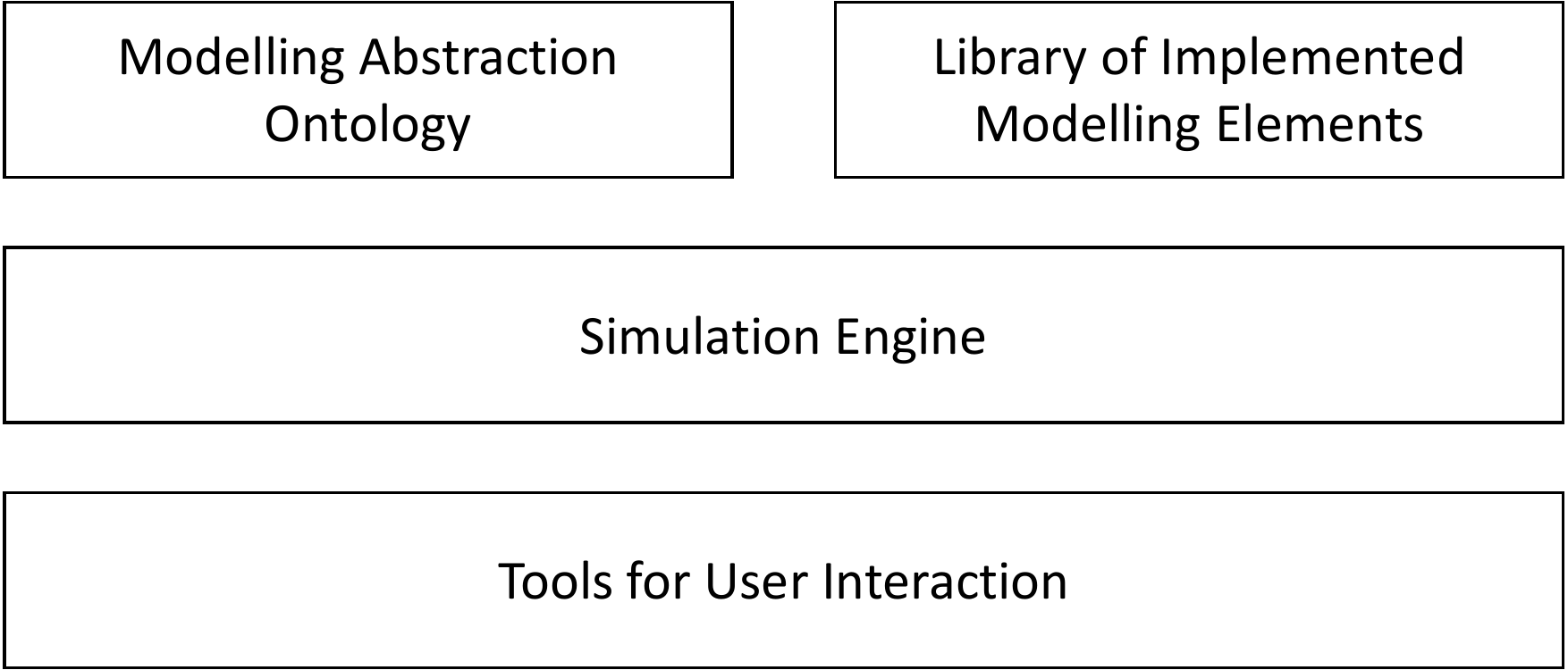}
	\caption{AgentPolis - Architecture (Simplified).}
	\label {fig:agentpolisArch}
\end{figure}

AgentPolis provides a set of abstractions, code libraries and software tools for building simulation models. The framework is structured in four main components (see Figure~\ref{fig:agentpolisArch}). The first component is described as the \textit{modelling abstraction ontology}. The theoretical concept of this component is to separate defined modelling abstractions from implementations of specific modelling elements. It uses an ontology in order to define more general concepts of multi-agent systems that result in a tailored structure for object-oriented programming when extending the simulation models for research-specific scenarios.  This allows for enforcement of implementations that consider interoperability of existing and additional research-specific modelling elements in their design. The second component is a \textit{library} of implemented modelling elements based on the given abstractions specified in the ontology. The library contains a set of modelling elements that represent common entities in transport systems. Apart from this, the third component can be described as the simulation engine. This component performs all calculations for running the simulation based on a discrete event model. Finally, the last component is a set of tools for user interaction, particularly for configuration and creation of the simulation model, data import, visualisation, etc. \newline
 
\begin{figure}[htbp]
	\centering
	\includegraphics[width=0.8\textwidth]{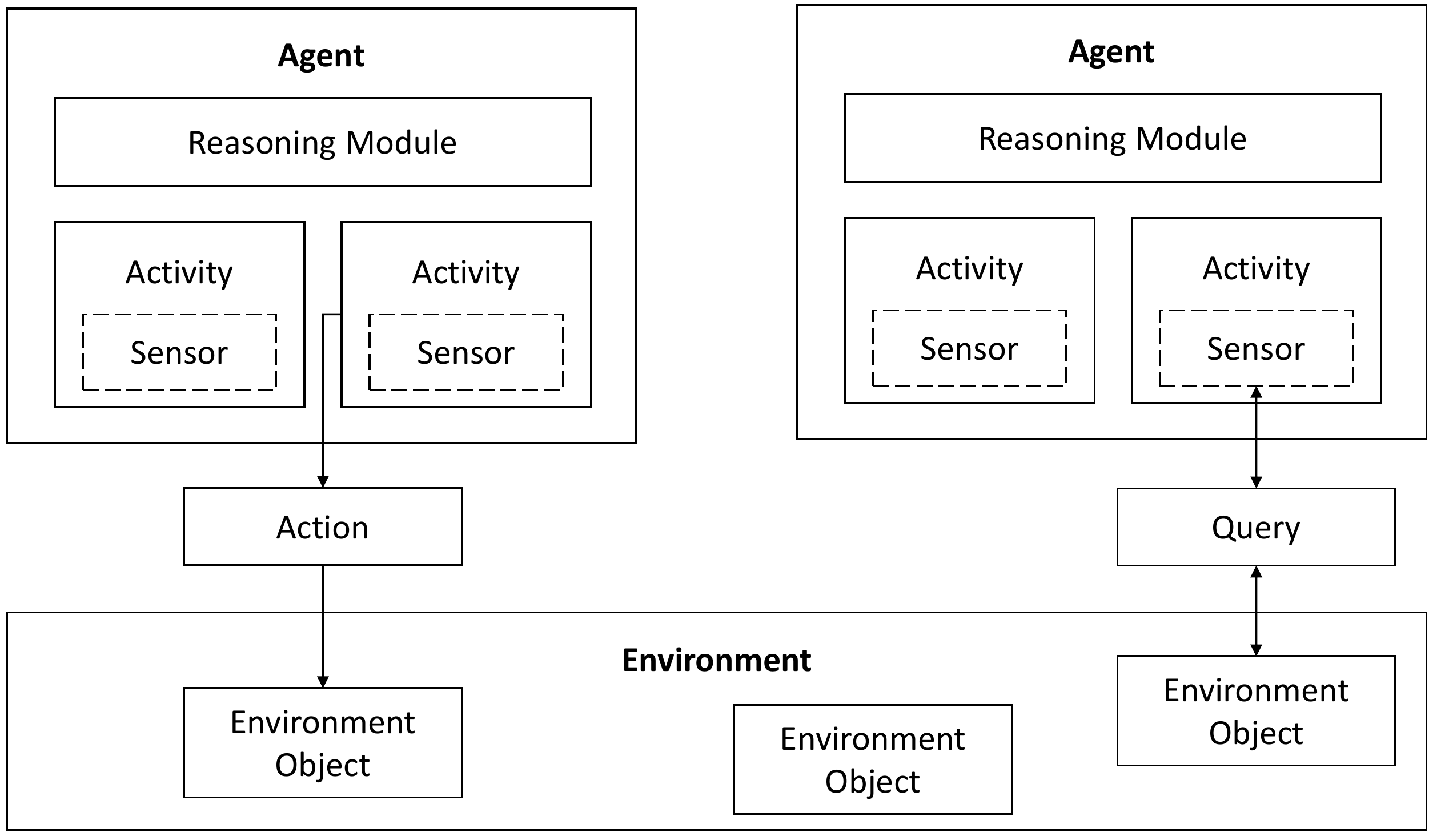}
	\caption{AgentPolis - Agent Structure.}
	\label {fig:agentpolis}
\end{figure}

With regard to the modelling capabilities, AgentPolis can be considered a mid-term simulator. Travel decisions refer to the level of route and modal choices. The agent structure is given by the abstraction ontology (see Figure~\ref{fig:agentpolis}) and defines concepts for the cognitive functions of the agent. Agents interact with objects in the environment using \textit{sensors} and \textit{activities}. Sensors perform queries to perceive environment objects while activities specify agent behaviour for initiating agent \textit{actions}. Agent actions model the effects of the agent on its environment e.g. a \textit{DriveVehicle} activity may result in a \textit{MoveVehicle} action. \cite{soares2013agent} mention a clear separation in modelling of driver decisions and vehicle control and therefore implements decision-making of activities in a separate \textit{reasoning module}. For this purpose, AgentPolis comes with implementation of a multimodal \textit{JourneyPlanner} based on a time-dependent graph \cite{hrnvcivr2013generalised}. \cite{jakob2012agentpolis} have extended AgentPolis with custom reasoning modules implementing different routing algorithms that were relevant to their experiments. AgentPolis follows an activity-based approach for modelling travel demand. The simulator includes a tool that generates an initial population of agents based on census data \cite{jakob2012agentpolis}. Based on the level of decision-making and implemented features, AgentPolis has been used and is suitable for simulating demand and decisions on the adoption of new mobility services.
\clearpage

\subsubsection{MovSim}
\label{sec:movsim}
MovSim \textit{(Multi-model open-source vehicular-traffic Simulator)} is an agent-based traffic simulator written in Java and licensed under GPLv3. The project started in the late 1990s at  TU Dresden and was designed for educational purposes \cite{treiber2010open}. In contrast to most available traffic simulation tools that model specific road networks (e.g. cities), MovSim focuses on the simulation of fundamental flow dynamics. For example, MovSim has been used to simulate the effects of  driver movements on traffic jams, studying the appearance of \textit{stop-and-go waves} \cite{kesting2008agents}. Because of this particular focus on flow dynamics, Movsim has also been applied for the simulation of rather unconventional scenarios such as ski marathons \cite{treiber2015drivers}. The simulator includes a number of reference implementations for established mathematical car-following models as described in~\cite{treiber2013traffic}. This can be relevant to simulate lane-changing and flow dynamics related to autonomous driving. \newline

\begin{figure}[h]
	\centering
	\includegraphics[width=0.85\textwidth]{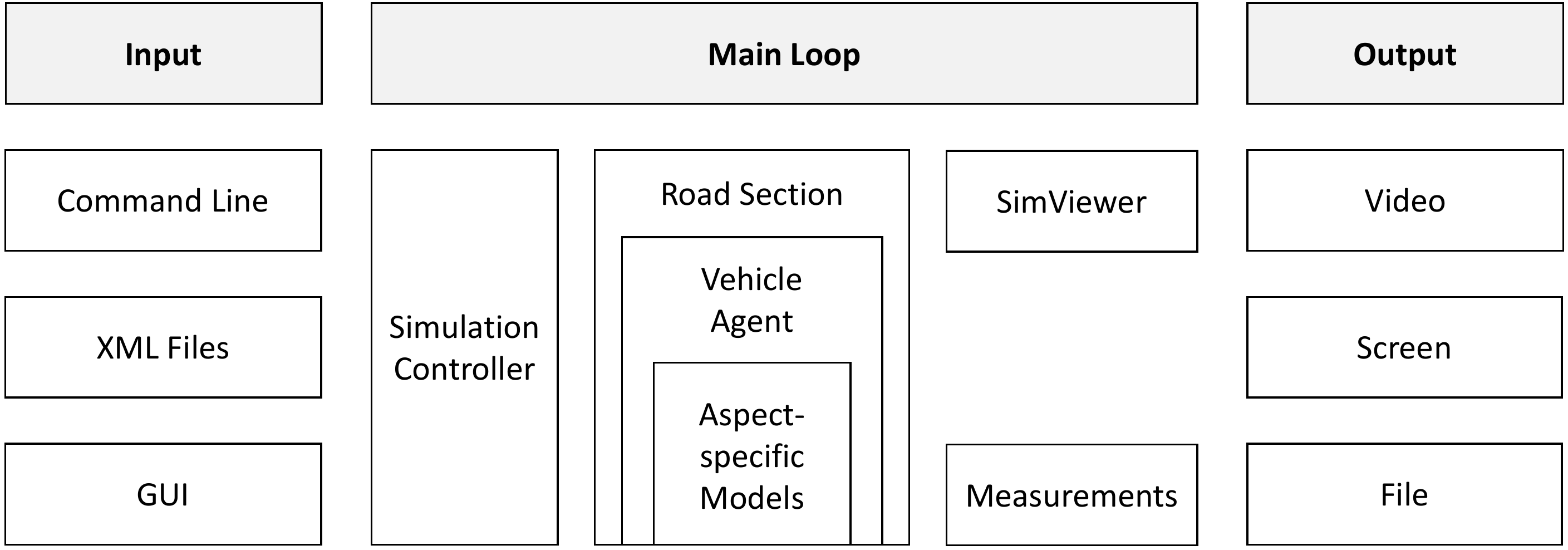}
	\caption{MovSim - Architecture.}
	\label {fig:movsim}
\end{figure}

The MovSim architecture is structured in three layers (see Figure \ref{fig:movsim}) \cite{kesting2008agents}. In the \textit{input layer}, simulation settings and parameters are defined. The user can input information either using a graphical user interface (GUI), command line or as XML files. This information is forwarded to the \textit{main loop layer}. In this layer, agent control and movement are implemented. The simulation controller continuously calculates the simulation in a loop as MovSim is based on a time-continuous model. The simulation controller primarily focuses on quantitative models. Different submodules implement logic for aspect-specific agent behaviour such as acceleration, braking, lane-changing, etc. 
Two additional modules act as observers to the simulation loop in order to extract information for the \textit{output layer}. The SimViewer module deals with information relevant for the visualisation of the simulated scenarios. MovSim includes implementation for both, 2D and 3D visualisation. Users can choose between a microscopic (cockpit perspective) or macroscopic (bird's eye) view of the simulation. If the visualisation is not used, simulated data can also be output as files. \newline

With regard to the modelling capabilities, MovSim can be considered a short-term simulator. Travel decisions refer to the level of agent movements such as acceleration, braking and lane changing. For this purpose, MovSim considers discrete-choice modelling. MovSim does not follow a trip- nor activity-based approach for demand modelling as route choices are irrelevant to the agent. Instead, agents represent particles in the network that move in space based on concepts of the applied car-following model. Hence, traffic volume can be defined using numerical input parameters.\footnote{see \url{www.traffic-simulation.de} - (access on 18/05/2020)} Given the short-term perspective in modelling movement-related driver decisions, MovSim can be useful to simulate flow behaviour in the field of autonomous driving. The integration of MovSim as a submodule of a larger simulation environment specifically for short-term aspects can be of interest.

\subsubsection{ATSim}
\label{sec:atsim}
ATSim (Agent-based Traffic Simulation System) is an application based on the commercial simulator AIMSUN, that extends AIMSUN \cite{barcelo2005dynamic} with agent capabilities. The project was first published in 2011 and has been developed at TU Clausthal \cite{chu2011atsim}. The authors of ATSim argue that for modelling the latest advances in  transportation, an agent-based approach is crucial to represent important aspects of modern transportation such as communication, goals and plans. However, existing agent-based simulators have not focused on an intuitive graphical user interface and exhibit a lack of tools for data collection and data analysis. This is why in the ATSim approach, the commercial simulator AIMSUN has been integrated with the JADE platform \cite{bellifemine2005jade}. This allows reuse of all features already implemented in AIMSUN while extending the simulator with agent capabilities. AIMSUN is used for modelling and simulation of traffic scenarios while implementation of agent behaviour is realised in JADE. ATSim has been used to simulate group-oriented traffic coordination in which groups of agents coordinate their speed and lane choices~\cite{gormer2012multiagent}. This can be relevant to simulate vehicle-to-vehicle (V2V) coordination dynamics related to autonomous driving.\newline 

\begin{figure}[htbp]
	\centering
	\includegraphics[width=0.9\textwidth]{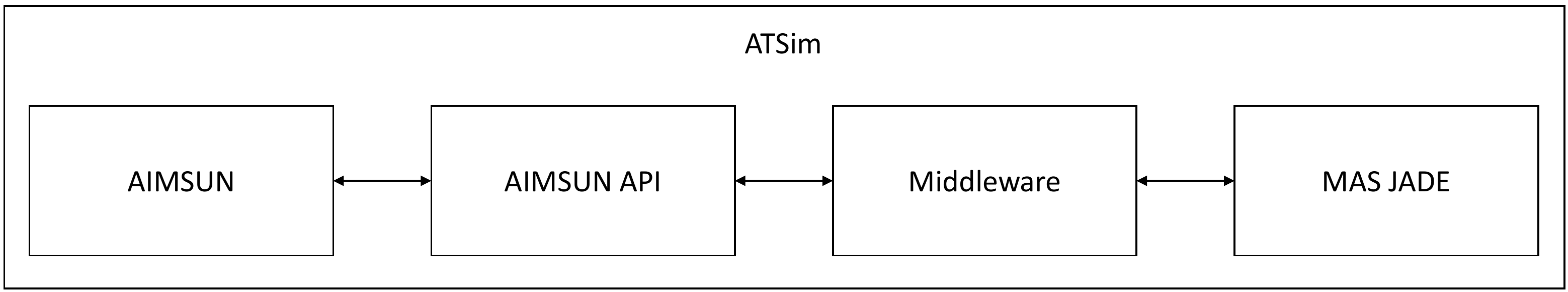}
	\caption{ATSim - Architecture.}
	\label {fig:atsim}
\end{figure}

The ATSim architecture is structured in four components (see Figure~\ref{fig:atsim}). The first component is the commercial AIMSUM simulator with all its features for modelling and simulating traffic scenarios. The second component is the multi-agent system based on JADE. This component is responsible for managing and controlling the agent life-cycle. In ATSim, agents are linked to various types of traffic objects in AIMSUN in order to extend AIMSUN objects with agent capabilities. Communication between agents and traffic objects is possible based on the AIMSUM API. AIMSUM provides an API for the integration of external services in Python and C++. However, JADE is based on Java and it is therefore necessary for ATSim to make use of a middleware in order to allow communication between AIMSUM and JADE. \newline

With regard to modelling capabilities, ATSim can be considered a mid- and short-term simulator. Travel decisions refer to the level route choice but also agent movements based on established car following and lane changing models implemented in AIMSUN. These models have been extended by agent capabilities for modelling perception and interaction of individual travellers. A distinction is made between \textit{static objects, objects with dynamic states} and \textit{mobile objects}. For example, the road network is represented as a static object whereas traffic lights are modelled as objects with dynamic states and vehicles are presented as mobile objects. Traffic objects can be assigned to an agent in JADE. Each agent can only control a single object in AIMSUM. The link between the agents and traffic objects is based on two assumptions. First, the agent life-cycle is synchronised with the life-cycle of the associated traffic object. Second, agents constantly receive updated information from the assigned traffic object after each simulation step. AIMSUM follows a trip-based approach for modelling travel demand using origin-destination matrices. The application has been used and thus is suitable for simulating V2V communication and coordination which is of growing relevance with the advancement of autonomous vehicles.

\section{Discussion}
\label{sec:discussion}
Based on the simulators reviewed (see Table \ref{tab:summarysimapps}), it is apparent that modelling of individuals deals with differing aspects depending on the area of application. In our review we used the time perspective to categorise the simulators by their capabilities to model individual behaviour. For example, long- and mid-term aspects are more relevant for examining research on resource utilisation, while the simulation of autonomous driving (new forms of mobility) has a greater focus on short- and mid-term behaviour. For researchers that need to simulate aspects from all three time perspectives, for example when examining the effects of autonomous mobility services on individuals migrating to (sub-)urban areas (long-term), their modal choices (mid-term) and effects of such services on flow dynamics and traffic safety (short-term), Simmobility is appropriate as it can model all three time scales. Based on earlier case studies of the simulator we have mentioned SimMobility as an example for research on resource utilisation. However, examples do not have to be used exclusively in the described context. For researchers that need a holistic approach to modelling of individual behaviour, SimMobility can be a suitable candidate for shortlisting. Otherwise, the decision remains scenario-specific.\newline

\begin{table}[h]
\centering
\caption{A summary of reviewed simulators.}
\label{tab:summarysimapps}
\resizebox{\textwidth}{!}{%
\begin{tabular}{|l|l|l|l|l|l|}
\hline
\textbf{\begin{tabular}[c]{@{}l@{}}Application \\ Name\end{tabular}} & \textbf{\begin{tabular}[c]{@{}l@{}}Area of \\ Application\end{tabular}} & \textbf{Licensing} & \textbf{\begin{tabular}[c]{@{}l@{}}Programming \\ Language\end{tabular}} & \textbf{Demand Modelling} & \textbf{\begin{tabular}[c]{@{}l@{}}Time Perspective on \\ Individual Behaviour\end{tabular}} \\ \hline
MATSim & \begin{tabular}[c]{@{}l@{}}Resource \\ Utilisation\end{tabular} & GPLv2 or later & Java & activity-based & mid-term \\ \hline
POLARIS & \begin{tabular}[c]{@{}l@{}}Resource \\ Utilisation\end{tabular} & \begin{tabular}[c]{@{}l@{}}Open-source\\ (license unclear)\end{tabular} & C++ & activity-based & mid-term \\ \hline
SimMobility & \begin{tabular}[c]{@{}l@{}}Resource \\ Utilisation\end{tabular} & \begin{tabular}[c]{@{}l@{}}SIMMOBILTITY Version \\ Control License \\ (see Github)\end{tabular} & C++ & activity-based & \begin{tabular}[c]{@{}l@{}}long-, mid- and \\ short-term\end{tabular} \\ \hline
\begin{tabular}[c]{@{}l@{}}SUMO + \\ JADE\end{tabular} & Connectivity & \begin{tabular}[c]{@{}l@{}}EPL 2.0 (SUMO)\\ LGPLv2 (JADE)\\ Apache 2.0 (TrasMAPI)\end{tabular} & C++, Java & \begin{tabular}[c]{@{}l@{}}activity-based or\\ trip-based using \\ OD matrices\end{tabular} & mid- and short-term \\ \hline
ITSUMO & Connectivity & \begin{tabular}[c]{@{}l@{}}Open-source\\ (license unclear)\end{tabular} & C++, Java & \begin{tabular}[c]{@{}l@{}}trip-based using \\ OD matrices of LSP\end{tabular} & mid-term \\ \hline
\begin{tabular}[c]{@{}l@{}}MATISSE\\ (DIVAs 4)\end{tabular} & Connectivity & GPLv3 & Java & trip-based using LSP & mid- and short-term \\ \hline
AgentPolis & \begin{tabular}[c]{@{}l@{}}New Forms of \\ Mobility\end{tabular} & GPLv3 & Java & \begin{tabular}[c]{@{}l@{}}trip-based using \\ OD matrices\end{tabular} & mid-term \\ \hline
MovSim & \begin{tabular}[c]{@{}l@{}}New Forms of \\ Mobility\end{tabular} & GPLv3 & Java & \begin{tabular}[c]{@{}l@{}}neither activity- nor trip-based.   \\ Only a numeric parameter to \\ specify number of travellers.\end{tabular} & short-term \\ \hline
ATSim & \begin{tabular}[c]{@{}l@{}}New Forms of \\ Mobility\end{tabular} & Commercial & \begin{tabular}[c]{@{}l@{}}C++, Python, \\ Java\end{tabular} & \begin{tabular}[c]{@{}l@{}}trip-based using \\ OD matrices\end{tabular} & mid- and short-term \\ \hline
\end{tabular}%
}
\end{table}

With regard to the first application domain, which deals with \textit{social dilemmas in resource utilisation} in the context of e-mobility, we consider SimMobility a more advanced approach in comparison to MATSim and POLARIS as it better handles the simulation of long-term aspects. This is particularly relevant to the simulation of urban areas as energy consumption is changing not only as a result of the electrification of vehicles, but also as a consequence of the increasing population caused by rural exodus. However, when dealing with mid-term scenarios MATSim and POLARIS can be just as powerful. In comparison to the other applications, MATSim probably has the largest user community and therefore is well documented whereas the POLARIS approach stands out in terms of the diversity of implemented features, as it combines various stand-alone applications into a single system.\newline

Regarding the second application domain, \textit{digital connectivity} benefits different aspects of the transportation system. When it comes to transportation safety, the MATISSE simulator is probably the most suitable application in this category as it specialises on this topic and provides dedicated features for modelling individuals in this context (e.g. driver distraction). However, for researchers that primarily want to test the effects of their custom algorithms in traffic management, ITSUMO can be more convenient as the application provides programming interfaces specifically for this purpose while already implementing a lot of details on individual behaviour (e.g. spontaneous or decentralised decision-making). The integration of SUMO and JADE can be relevant when used in the context of an ATS to compare different simulation models in a virtual environment.\newline

The last application domain, \textit{new forms of mobility}, is similarly diverse. When assessing the adoption of new mobility services, AgentPolis can be a good choice as the application focuses on the aspect of interaction when modelling individuals. This is particularly relevant as the growing portfolio of mobility services and continuous access to real-time information via smartphones have led to this dynamic. However, when dealing with research on autonomous driving modelling of individual behaviour focuses on movement related aspects. For this purpose, MovSim and ATSim can be of interest. MovSim in comparison to ATSim is a more lightweight simulator that exclusively deals with movement-related driving decisions from a theoretical perspective. ATSim can be used for the same type of decisions but is applied on real-world networks and includes mid-term decisions such as route choice. \newline

The current state of implementation for modelling of individuals already shows a broad spectrum of features depending on area of application. Individual behaviour is modelled by traffic-related decisions at different levels of detail, e.g. lane changing vs. route or modal choices. However, we noticed that available simulators have focused on simulating traffic as the primary subject and thus leave scenario-specific aspects to the responsibility of end-users. Initially, a focus on traffic-related modelling aspects appears obvious as platform developers cannot anticipate the full range of scenarios for which their simulators will eventually be used. Following the same line of reasoning, developers need to assume that their applications will eventually be customised to fit specific research purposes. It is therefore desirable that common and foreseeable modifications are supported by suitable structures and programming interfaces. With regard to the modelling of individual behaviour, it is important to align traveller decisions with the context of the simulation. Traveller decisions are based on individual preferences and personal objectives. Furthermore, travel behaviour is typically driven by a purpose which plays a crucial role in the individual's perception of personal preferences. For example, \textit{time/punctuality} has a different value when commuting to work as compared to a social visit. However, in the current state of implementation there is a lack of concepts to capture these preferences and objectives as determining factors of individual decisions. This hampers customisation, especially for interdisciplinary researchers that are not thoroughly experienced with the simulators. Implementations that elaborate on modelling of these aspects can help reduce customisation efforts and thus attract more researchers to make use of available simulators rather than developing individual solutions as described by \cite{jakob2013modular}.

\section{Conclusion}
As diverse as the spectrum of research questions found in mobility, there is also a great variety of simulators that focus on different aspects of the transportation system and are differing in their underlying methods. Thus, getting an overview and selecting a suitable simulator can be time-consuming, involving a lot of in-depth research. The success of new ideas to solve current issues of transportation, such as sharing services or autonomous driving, relies on the acceptance and behaviour of individuals. Hence, it is important to focus on the individual when dealing with current research on mobility.
For this purpose, computer-based simulations are an established means. The application of agent technology is particularly suitable to investigate road traffic from the individual perspective, as it allows for modelling of individuals with intelligent and autonomous behaviour. Current state of implementation includes a broad spectrum of features for modelling of individuals. Features are linked to the area of application, modelling individual behaviour at different levels of detail. Based on current research topics in the field of mobility, we have reviewed example simulators with related case studies and discussed the suitability of these simulators for specific research purposes. In particular, we have looked at the capabilities of the simulators to model individuals and their travel behaviour. Travel behaviour typically is linked to the context of the simulation and therefore needs to be adjusted. Currently, there is a lack of concepts to support this type of adjustments in the simulators, which hampers customisation especially for interdisciplinary researchers that are not thoroughly experienced with the simulators. Implementation for these types of adjustments can help to attract more researchers that deal with individual behaviour in mobility to make use of available simulators. 

\section{Acknowledgement}
This research has been supported by a grant from the Karl-Vossloh-Stiftung (Project Number S0047/10053/2019).

\bibliographystyle{unsrt}  
\bibliography{Bibliography} 


\end{document}